# Localization of nanoscale objects with light singularities

Thomas A. Grant[1], Anton N. Vetlugin[2], Eric Plum[1], Kevin F. MacDonald[1], and Nikolay I. Zheludev[1,2]

[1] *Optoelectronics Research Centre, University of Southampton, SO17 1BJ, UK*

[2] *Centre for Disruptive Photonic Technologies, SPMS & TPI, Nanyang Technological University, Singapore, 637371, Singapore*

Unprecedented atomic-scale measurement resolution has recently been demonstrated in single-shot optical localization metrology based on deep-learning analyses of diffraction patterns of topologically structured light scattered from objects. Here we show that variations in the diffraction patterns caused by positional changes of an object depend upon the spatial derivatives of the magnitude and phase of the incident field, with the latter strongly enhanced at phase singularities. Despite lower intensity near the singularity, an orders-of-magnitude increase in Fisher information contained in the diffraction patterns can be achieved when a nano-object is illuminated by light containing phase singularities, rather than a plane wave. Our work provides a fundamental explanation and motivation for singularity-based metrology with deeply subwavelength precision.

In recent decades, progress in optical super-resolution microscopy and metrology has been driven by nonlinear and statistical techniques [1-11], structured illumination microscopy [12-14], and computational imaging techniques for retrieving phase from scattered light [15-21], often taking advantage of object sparsity [22-24]. The ability of neural networks to efficiently solve the inverse scattering problem has also been demonstrated [25], and superoscillatory (topologically structured) light fields have lately been applied to microscopy and optical metrological applications in a manner similar to computational imaging.

The phenomenon of optical superoscillation was first introduced [26] in 2006 and experimentally identified shortly thereafter [27]. It describes rapid subwavelength spatial variations of intensity and phase in complex electromagnetic fields formed by the interference of several coherent waves, and its discovery stimulated a significant revision of the limits of classical electromagnetism. In particular, the study of the structure of superoscillatory fields in free space revealed arbitrarily small energy 'hotspots', facilitated by the presence of phase singularities bordering regions of energy backflow [28, 29]. Such regions, and the associated phase singularities, can be orders of magnitude smaller than the wavelength, implying that their interaction with matter should vary on similarly short, subwavelength scales making their application an intriguing prospect for metrology.

Berry and Nye proposed a form of singularity-based metrology in the 1970s, suggesting that singularities (referred to then as 'wave dislocations') in radio pulses reflected by the rock bed of a glacier could be employed as subwavelength markers for echo-sounding-based depth measurements [30, 31]. More recently, dimensional and positional measurements with deeply subwavelength resolution have been achieved via deep learning analysis of objects' diffraction patterns [32-34]. With topologically structured illumination and 'in-situ' neural network training, such measurements can localize the average position of a nanowire with precision and accuracy down to ~100 pm using visible light [35, 36], beating the diffraction limit of conventional optical instruments thousands of times over.

In this work, we mathematically describe and numerically demonstrate that the scattering from an object located near a singularity in a topologically structured field has higher information content than the scattered field from a plane wave, thereby enabling greater precision and accuracy of metrology upon its analysis.

For simplicity, we consider a two-dimensional implementation of the Rayleigh-Sommerfeld model of diffraction - a mathematical manifestation of the Huygens-Fresnel principle [37]. In this case, the scattered field is expressed as a superposition of diverging circular waves radiating from a 1D array of points describing the scattering object (along $x$ at $z = 0$),

$$U(x) \propto i \int \tilde{U}(x') \frac{\exp(i2\pi r/\lambda)}{r} \cos\theta \, dx'$$

where $\tilde{U}(x')$ and $U(x)$ denote the complex field of a monochromatic wave, with wavelength $\lambda$, at the object and the detector respectively; which are separated by a distance $h$ in the propagation direction $z$, whereby $r = \sqrt{(x-x')^2 + h^2}$, and $\theta = \arctan\left(\frac{x-x'}{h}\right)$.

As an archetypal scattering object, we consider a narrow slit in an otherwise opaque screen, with edges located at $x' = a \pm \delta$ (i.e. a slit of width $2\delta$ centred at $x' = a$). We assume that a complex field, $\tilde{U}(x') = A(x')e^{i\phi(x')}$, is normally incident on the screen and is transmitted only through the slit. Following an integration by parts, the scattered field, $U(x)$ at the detector can be written as a sum of three contributions:

$$U(x) = U_1(x) + U_2(x) + U_3(x)$$

where

$$U_1(x) = A(a+\delta)e^{i\phi(a+\delta)} \xi(x, a+\delta) - A(a-\delta)e^{i\phi(a-\delta)} \xi(x, a-\delta)$$

$$U_2(x) = -\int_{a-\delta}^{a+\delta} \frac{dA(x')}{dx'} e^{i\phi(x')} \xi(x, x') dx'$$

$$U_3(x) = -i \int_{a-\delta}^{a+\delta} \frac{d\phi(x')}{dx'} A(x') e^{i\phi(x')} \xi(x, x') dx'$$

and

$$\xi(x, x') \propto i \int \frac{\exp(i2\pi r/\lambda)}{r} \cos\theta \, dx'.$$

Here, $U_1$ is the only term present in the diffracted field from an incident plane wave, while $U_2$ and $U_3$ are respectively dependent on variations of the amplitude and phase of the incident field. The changes in these additional contributions to the scattered field for a structured incident field, arising from changes in the object plane, can become significant in comparison to the associated change in $U_1$. Thus, the spatially fast-changing features of a structured incident field can cause changes in $U_2$ and $U_3$ to dominate the total change in the scattered field.

As a practically relevant example, following Refs. [38-40], we now consider a superoscillatory field, formed by the linear combination of two band-limited, prolate spheroidal wave functions (PSWFs): $\tilde{U}(x') = [21.65 S_2(x') + S_3(x')]W$, with $W = 0.00021$. While the two individual PSWFs are band-limited to $|k_0| = \omega/c$, $\tilde{U}(x')$ has a central peak focused beyond this limit (a full-width at half-maximum of $0.3\lambda$), flanked by a series of phase singularities (Fig. 1).

As detailed in Ref. [41], the phase and amplitude mask required to generate this superoscillatory field from a plane wave can be obtained by transforming the required object-plane field $\tilde{U}(x')$ into a Fourier series (PSWFs being eigenfunctions of a finite, band-limited Fourier transform); backpropagating to the desired mask plane; and then executing an inverse Fourier transform. Here, we assume a mask plane at a distance $d = 30\lambda$ from the object plane, under which condition the intensity at the peak of the superoscillatory field's central hotspot $U(x')U(x')^*$ is approximately twice (2.06×) the intensity of the plane wave incident upon the mask, $U_0 U_0^*$.

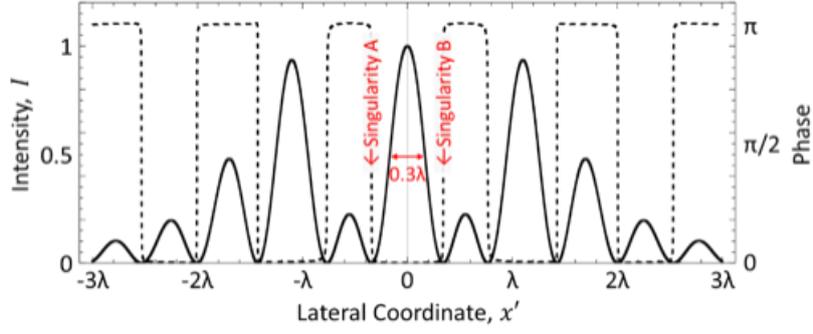

**Fig. 1:** Superoscillatory field profile. Intensity $I(x') = \tilde{U}(x')\tilde{U}(x')^*$ [solid line], and corresponding phase $\phi(x')$ [dashed line] profiles, of the superoscillatory field $\tilde{U}(x') = [21.65 S_2(x') + S_3(x')]W$.

As a target object, we consider a slit of width $2\delta = \lambda/10$ in an opaque film (Fig. 2). We assume that measurements are performed by analysing its scattering pattern in an imaging plane located at a distance $h = 4\lambda$ from the slit. From a practical perspective, the image sensor (detector) does not have to be at the imaging plane: the scattered field at this point is formed of free-space propagating waves, so it can be transformed to the detector plane by a conventional lens at any magnification, without loss of resolution (as has been shown experimentally [35, 42]). In what follows, we assume an imaging plane detection aperture at $-\frac{L}{2} < x < \frac{L}{2}$, where $L = 12\lambda$ (i.e. $\gg \delta, a$).

To quantify and compare the effectiveness of localization metrology with different incident fields, we adopt the Fisher information metric, a metric that quantifies the amount of the information about a known parameter, contained within a measurement [43]. For example, in quantum metrology Fisher information can be used to derive the fundamental limit in parameter retrieval problems such as resolving incoherent point sources [44], time-varying waveform estimation [45], and quantum imaging [46], among others [47]. More recently, the Fisher information in classical optical scattering problems has been studied [48, 49], and proposed as an optimization tool for scattering-based parameter estimation problems [50-52].

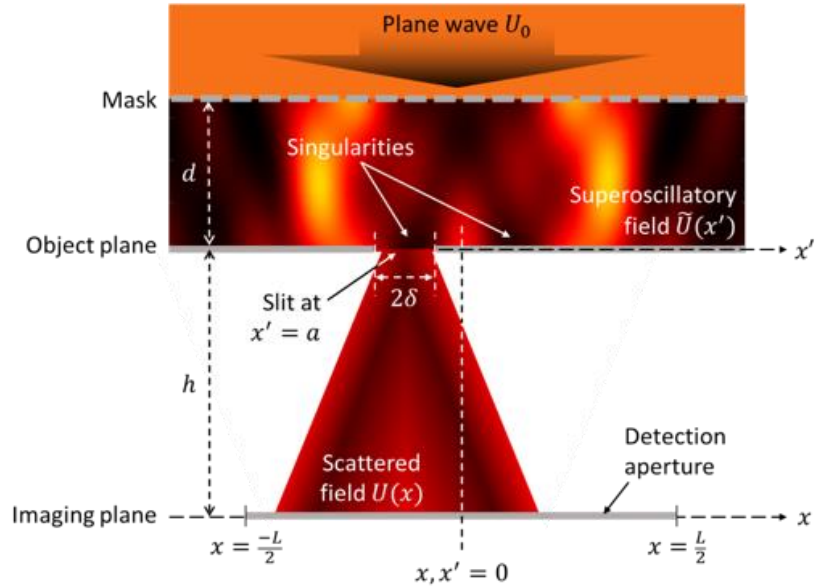

**Fig. 2:** Scattering of a topologically structured field by a slit in an opaque screen. A plane wave $U_0$ is incident upon a phase and amplitude mask, which generates a superoscillatory field $\tilde{U}(x')$ in the object plane $x'$, with a central hotspot located at $x' = 0$. A slit of width $2\delta$ is centred at $x' = a$ in the object plane. The scattered intensity $I(x) = U(x)U(x)^*$ is measured in the imaging plane $x$, at a distance $h = 4\lambda$ beyond the object plane, with the imaging plane section $(-L)/2 < x < L/2$ being projected onto a detector.

In the present case, we calculate the Fisher information by treating the scattered field as a photodetection probability distribution function, for which we use $p(x; a)$ - the normalized distribution of the detected power (as a function of position $x$ on the detector) at the image plane, for a given slit position, $a$:

$$p(x; a) = \frac{U(x, a)U(x, a)^*}{\int_{-L/2}^{+L/2} U(x, a)U(x, a)^* dx}$$

From the measurement perspective, this density function and the relative rate of change in its log-likelihood function is important. The latter is known as the score function, and its weighted square, integrated over the detection range $L$, is a figure of merit for how rapidly the profile of the scattered field changes in response to a change in the position $a$ of the slit:

$$F(a) = \int_{-\frac{L}{2}}^{+\frac{L}{2}} \left[\frac{\partial}{\partial a} \ln p(x; a)\right]^2 p(x; a) dx$$

This can be considered the information per photodetection, such that the total Fisher information content within the scattered field is derived by scaling for the total power detected:

$$F_{TOT}(a) \sim F(a) \int_{-\frac{L}{2}}^{+\frac{L}{2}} U(x, a)U(x, a)^* dx$$

Figure 3 shows that the Fisher information content of a scattered superoscillatory field depends strongly on the position of the scattering object within the incident field – in this case, most prominently on the position of the slit relative to the phase singularities (as opposed to the central intensity hotspot). Note that there is no dependence of Fisher information on slit position for an incident plane wave, because while the position of the diffraction pattern in the imaging plane shifts with the slit position in the object plane, its intensity profile is invariant. For the superoscillatory field, the Fisher information increases sharply, peaking at $a = \pm 0.34\lambda$, when either singularity A or B (on either side of the incident field's central hotspot – see Fig. 1) is near the center of the slit. The double- or split-peak structure, shown in the Fig. 3a inset, is the result of a saddle point in the profile of the scattered field when the slit is perfectly centered on the singularity, whereby the profile is slightly less sensitive to changes in slit position than for off-center alignments. At the maxima, the intensity profile of the diffraction pattern changes rapidly as a function of $a$, yielding a $10^6$-fold enhancement in Fisher information (Fig. 3a), as compared to the plane wave from which the superoscillatory field was generated.

The fact that incident (and therefore scattered) intensity is exponentially lower in the vicinity of phase singularities, relative to the case of a plane wave, must be considered in this comparison: Fig. 3b demonstrates that using a topologically structured incident field – i.e. probing the target object with an incident field containing singularities – nonetheless provides significant advantage, with the total information content in the detected scattered field being enhanced by a factor of ~250, again as compared to the plane wave from which the superoscillatory field was generated.

It is also important to account for the fact that measurements at low intensities near a singularity are more susceptible to noise. Within the framework of Fisher information, this can be considered as follows: Each scattered field measurement (photodetection event) provides information $F(a)$, while (detector) noise-related photodetection events provide zero information. Fisher information is additive, so the reduction of information due to the presence of noise is proportional simply to the ratio of scattered field to noise detection events. In terms of intensity:

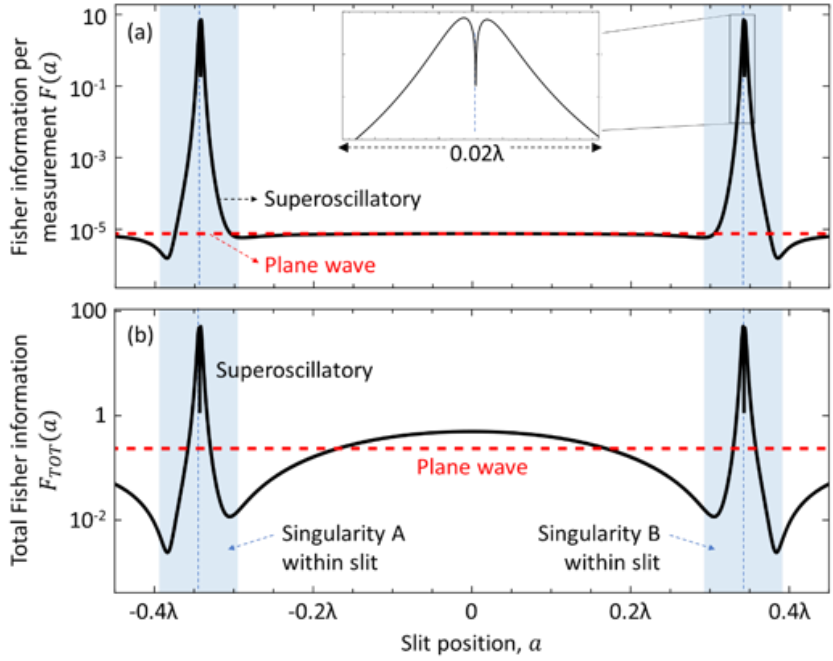

**Fig. 3:** Fisher information content of the field scattered by a subwavelength slit. (a) Fisher information per measurement and (b) total Fisher information as functions of slit position for: a superoscillatory incident field with amplitude $[21.65 S_2(x') + S_3(x')]W$ – shown as solid black lines; a plane wave incident field with an intensity equal to half that of the superoscillatory hotspot – shown as dashed red lines. The blue shaded bands denote the range of positions over which a singularity (A or B, as labelled in Fig. 1) is located within the slit.

$$F(a, I_{noise}) = \int_{-\frac{L}{2}}^{+\frac{L}{2}} \left[\frac{d}{da}\ln(p(x;a))\right]^2 p(x;a) \frac{I(x;a)}{I(x;a) + I_{noise}} dx$$

Figure 4 shows total Fisher information as a function of signal-to-noise ratio (SNR). For consistent comparison, we assume the same plane wave intensity as used for generation of the superoscillatory field, and the same level of absolute noise in both cases. At high SNR (>1000), the advantage of the superoscillatory incident field is obvious: Fisher information is orders of magnitude higher than for a plane wave. With decreasing SNR, the information content of the scattered field falls faster for the superoscillatory field, and its metrological advantage disappears at signal to noise ratios <50.

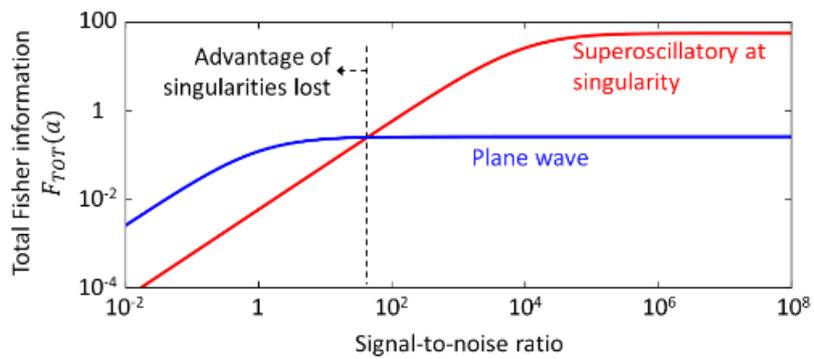

**Fig. 4:** Total Fisher information as a function of signal-to-noise ratio for a superoscillatory incident field (assuming the scattering to be optimally located at a singularity) [red line] and for a plane wave [blue line].

In summary, this study provides a fundamental explanation, and justification, for singularity-based metrology. We demonstrate that when probing a nanoscale object, a significant advantage can be gained from exploiting phase singularities in a topologically structured incident field, as compared to plane wave illumination. Despite the low intensity in the vicinity of singularities, and in the presence of detector noise, the Fisher information content of a nano-object's scattering pattern can be orders of magnitude larger when it is illuminated with a topologically structured field with phase singularities, as opposed to a plane wave. We show analytically that this advantage – seen experimentally in the form of enhanced measurement precision and accuracy [35, 36] – is derived from the strong dependence of scattered intensity profile on local intensity and phase gradients in the incident field at the object plane: small, deeply subwavelength changes in the position of a scattering object relative to a singularity, can lead to large changes in the scattered field. The method presented here provides a framework for optimization of the incident intensity and phase profile and object-to-imaging plane distance depending on the shape class and size range of objects.

This work was supported by the Engineering and Physical Sciences Research Council, UK (grant number EP/T02643X/1), and the National Research Foundation, Singapore (NRF-CRP23-2019-0006).